\documentstyle[12pt]{article}
\input epsf
\def\be{\begin{equation}}
\def\ee{\end{equation}}
\def\bea{\begin{eqnarray}}
\def\eea{\end{eqnarray}}

\def\br{\begin{thebibliography}{abc}}
\def\er{\end{thebibliography}}

\def\Hat#1{\rlap{\kern.10em$\widehat{\phantom G}$}#1}
\def\HAt#1{\rlap{\kern.05em$\widehat{\phantom G}$}#1}

\def\czp#1{\rlap{\kern.1em$\widehat{\phantom{G\vrule height.8em}}$}#1{}}
\def\Czp#1{\rlap{\kern.05em$\widehat{\phantom{G\vrule height.8em}}$}#1{}}

\newcommand{\sect}[1]{\setcounter{equation}{0}\section{#1}}

\def\sxn#1{\bigskip\medskip \sect{#1} \smallskip
                                                 }

\begin{document}

\setcounter{page}{0}
\begin{flushright}
Napoli: DSF-T-5/95\\
Syracuse: SU-4240-604\\
Trieste: IC/95/24\\
March 1995\\
\end{flushright}
\vskip 1.5cm\begin{center}
{\LARGE \bf TOPOLOGY CHANGE AND}
\vskip 0.2cm
{\LARGE \bf QUANTUM PHYSICS}
\end{center}
\vskip 1.0cm
\centerline {\large A. P. Balachandran$^{1}$, G. Bimonte$^{2,3}$,
			G. Marmo$^{3,4}$}
\centerline {\large    and A. Simoni$^{3,4}$}
\vspace{5mm}
\centerline {\it $^1$ Department of Physics, Syracuse University,
Syracuse, NY 13244-1130.}
\vspace{2.25mm}
\centerline {\it $^2$ International Centre for Theoretical Physics,\\
P.O. Box 586, I-34100, Trieste, Italy.}
\vspace{2.25mm}
\centerline {\it $^3$ INFN, Sezione di Napoli, Napoli, Italy.}
\vspace{2.25mm}
\centerline {\it $^4$ Dipartimento di Scienze Fisiche, Universit\`a di
Napoli,}
\centerline{\it Mostra d' Oltremare, Pad. 19, I-80125, Napoli, Italy.}

\vspace{.5cm}
\begin{abstract}
The role of topology in elementary quantum physics is discussed in
detail. It is argued that attributes of classical spatial topology emerge from
properties of state vectors with suitably smooth time evolution.
Equivalently, they emerge from considerations on the domain of the
quantum Hamiltonian, this domain being often specified by boundary
conditions in elementary quantum physics. Several examples are presented
where classical topology is changed by smoothly altering the boundary
conditions. When the parameters labelling the latter are treated as
quantum variables, quantum states need not give a well-defined classical
 topology,
instead they can give  a quantum superposition of such topologies. An existing
argument of Sorkin based on the spin-statistics connection and indicating
the necessity of topology change in quantum gravity is recalled. It is
suggested therefrom and our results here that Einstein gravity and its minor
variants are
effective theories of a deeper description with additional novel degrees of
freedom. Other reasons for suspecting such a microstructure are also
summarized.
\end{abstract}

\newpage
\setcounter{footnote}{1}

\setcounter{page}{1}

\sxn{Introduction}\label{se:1}

There are indications from theoretical considerations that spatial
topology in quantum gravity can not be a time-invariant attribute, and
that its transmutations must be permitted in any eventual theory.

The best evidence for the necessity of topology change comes from the
examination of the spin-statistics connection for the so-called geons
\cite{fri,sor,bal1}. Geons are solitonic excitations caused by twists in
spatial topology. In the absence of topology change, a geon can neither
annihilate nor be pair produced with a partner geon, so that no geon has
an associated antigeon.

Now spin-statistics theorems generally emerge in theories admitting
creation-annihilation processes \cite{bal2,dow}. It can therefore be
expected to fail for geons in gravity theories with no topology change.
Calculations on geon quantization in fact confirm this expectation
\cite{sor,bal2,dow}.

The absence of a universal spin-statistics connection in these gravity
theories is much like its absence for a conventional nonrelativistic
quantum particle which too cannot be pair produced or annihilated.
Such a particle can obey any sort of statistics including parastatistics
regardless of its intrinsic spin. But the standard  spin-statistics
connection can be enforced in nonrelativistic dynamics also by
introducing suitable creation-annihilation processes \cite{bal2}.

There is now a general opinion that the spin-statistics theorem should
extend to gravity as well. Just as this theorem emerges from even
nonrelativistic physics once it admits pair production and
annihilation \cite{bal2},
quantum gravity too can be expected to become compatible with this
theorem after it allows suitable topology change
\cite{see}. In this
manner, the desire for the usual spin-statistics connection leads us to
look for a quantum gravity with transmuting topology.

Canonical quantum gravity in its elementary form is predicated on the
hypothesis that spacetime topology is of the form $\Sigma \times {\bf R}$
(with
${\bf R}$ accounting for time ) and has an eternal spatial topology. This
fact
has led to numerous suggestions that conventional canonical gravity is
inadequate if not wrong, and must be circumvented by radical revisions of
spacetime concepts \cite{bom}, or by improved approaches based either on
functional integrals and cobordism or on alternative quantisation methods.

Ideas on topology change were first articulated in quantum gravity, and
more specifically in attempts at semiclassical quantisation of classical
gravity. Also it is an attribute intimately linked to
gravity in the physicist's  mind. These connections and the
apparently revolutionary nature of topology change as an idea have led to
extravagant speculations about twinkling topology in quantum gravity and
their impact on fundamental conceptions in physics.

In this paper, we wish to point out that models of quantum particles
exist which admit topology change or contain states with no well-defined
classical topology.{\em This is so even though gravity does not have a central
role in our ideas
[some of which were previously reported in refs. \cite{bal3} and
\cite{bal4}]
and is significant only to the extent that metric is important for a
matter Hamiltonian.} These models  use only known physical
principles and have no revolutionary content, and at least suggest that
topology change in quantum gravity too may be achieved with a modest
physical input and no drastic alteration of basic laws.

We explain our views \cite{bal3,bal4} regarding the role of spatial
topology in elementary quantum physics in Section 2, and in particular
emphasize that domains of observables \cite{bal5} and smoothness of time
evolution have much to say on this matter. Section 3 develops these
observations for a particle which moves on the union of two intervals.
The domain of its Hamiltonian is characterized by an element $u$ of
$U(2)$ under the simplifying demand that momentum too be (essentially)
self-adjoint on that domain. It is then the case that continuity
properties of probability densities are compatible with continuous
functions on two circles ($S^1 \bigcup S^1$) for certain $u$'s and with a
single circle for certain other $u$'s. The configuration space of the
particle is thus $S^1 \bigcup S^1$ and $S^1$ for these two $u$, whereas
it is just two intervals for the remaining $u$.
We then argue that topology
change  can be achieved by looking upon $u$ as an external parameter and
continuously changing it from one value to another.

It is not quite satisfactory to have to regard $u$ as an external
parameter and not subject it to quantum rules . In Section 4, we therefore
promote it to an operator, introduce its conjugate variables and modify
the Hamiltonian as well to account for its dynamics. The result is a
closed quantum system. It has no  state with a sharply defined
$u$. We cannot therefore associate one or two circles with the quantum
particle and quantum spatial topology has to be regarded as a superposition of
classical spatial topologies. Depending on our choice of the Hamiltonian, it is
possible to prepare states where topology is peaked at one or two $S^1$'s
for a long time, or arrange matters so that there is transmutation from
one of these states to the other.

Section 5 generalizes these
considerations to higher dimensions and establishes that similar effects
can be achieved in all dimensions  by manipulating boundary conditions
and their dynamics \cite{imb}.

It is the contention of this paper that topology change can be achieved
already in elementary quantum physics. We realise this phenomenon by
promoting parameters entering boundary conditions to control parameters or
degrees of freedom, in such a manner that the states of the latter affect
 spatial topology. There is a close relation of these ideas to what
happens in the axion solution to the strong CP-problem \cite{axi}, as we
explain in the concluding Section 6. All this suggests that topology
change is facilitated by the addition of degrees of freedom.

We have remarked on the desirability of topology change from the point of
view of the spin-statistics theorem. There is perhaps a hint here that
{\em quantum gravity with topology change has novel degrees of freedom and not
just those of Einstein gravity or its minor variants.} We summarize
further evidence for this point of view, taken also from geon
physics, in Section 6. It is a matter for regret that all such ideas on quantum
gravity must for now remain beyond experimental control.

\sxn{Topology in Quantum Physics}\label{se:2}

Quantum systems with classical attributes are generally characterized by
infinite dimensional Hilbert spaces. This is the case already for
elementary systems such as that of a particle on a manifold.

For a system like this, it is never the case that we can realistically
observe all self-adjoint operators with equal ease. For example, the
eigenstates of many of these operators  have infinite mean values for
energy, its square or angular momentum. In conventional quantum physics,
they are tacitly  discarded as observables for the simple reason that
their measurement is tantamount to the preparation of the above
unphysical eigenstates.

We thus see that a self-adjoint operator can be an observable only if it
has additional attributes. Those which circumvent the problem of
unphysical eigenstates can be formulated
using considerations on domains
\cite{bal5}. A simple formulation of these attributes can be achieved if it is
agreed that
time evolution, and hence the Hamiltonian generating it, have very
special roles in physics. It goes as follows.

Recall that the Hamiltonian $H$ generally is an unbounded operator and
cannot be applied on all vectors of the quantum physical Hilbert space
${\cal H}$. Rather it can be applied only on vectors of its domain $D(H)$
[10]. The latter is dense in ${\cal H}$, but is not all of ${\cal H}$,
and is often specified by boundary conditions in simple quantum systems.
[We will see examples of domains in subsequent sections.] The attribute in
question of any observable ${\cal O}$, having only discrete spectrum, is
that
{\em it is a self-adjoint operator having all its eigenvectors in}~
$D(H)$.

If ${\cal O}$ has (also) continuous spectrum, the definition of the
conditions under which it is observable requires a little elaboration.
Let us notice first that we  cannot
prepare a quantum state by observing a point of its continuous spectrum
with no
experimental uncertainty at all,
so that
the above attribute cannot be used now.

The conditions under which such an ${\cal O}$ is an observable is facilitated
by first considering a physical state
$\psi$ that has already  been prepared by means of a complete set of
measurements of observables having only discrete spectra. The previous
definition of observables then implies that $\psi \in D(H)$. Suppose now
that a measurement of $\cal O$ is performed on $\psi$ and that we observe
the value $x$ for $\cal O$ ($x$ belonging to the spectrum
$S({\cal O})$ of ${\cal O}$
)
with an experimental uncertainty $\epsilon$. We can  think of associating
 the set
$\Omega_{x,\epsilon}=[x-\epsilon, \, x+\epsilon] \bigcap S({\cal O})$ with this
result.
Then we know from the general
principles of quantum theory that an instant after the measurement, the
state of the system will be given by the new vector
$\chi_{x,\epsilon}({\cal O})\psi$, where
$\chi_{x,\epsilon}(y)$ is the characteristic function of the interval
$[x-\epsilon,\, x+\epsilon]$. (For the definition of bounded functions of
self-adjoint operators, see \cite{rs}). Now, before going further, we realize
at once that even this is an idealization not corresponding to reality,
for the boundaries of $\Omega_{x,\epsilon}$ cannot be specified with
absolute precision. We thus replace the characteristic function
$\chi_{x,\epsilon}(y)$ with some smooth, real function $f^{\infty}_
{x,\epsilon}$, of fast
decrease, approximately supported in $\Omega_{x,\epsilon}$.  Summing up, our
criterion for $\cal O$ to be an
observable is that, for any such $f^{\infty}_{x,\epsilon}$, the
operator $f^{\infty}_{x,\epsilon}({\cal O})$ should leave $D(H)$
invariant.
This definition guarantees that also after the measurement the state
of the system will be in the domain of the Hamiltonian.

We will hereafter accept these properties as fundamental attributes of
any observable. The domain $D(H)$ in our scheme thus has a central role
in quantum physics.

This domain strongly reflects the properties of the classical configuration
space $Q$: In conventional quantum physics we generally insist for
example that the density function $\psi ^* \chi$ for any pair of vectors
$\psi, \chi \in D(H)$ is a continuous function on $Q$, $\psi^*\chi \in
C(Q)$.  Conversely, $Q$ can be recovered as a
topological space from the  $C^*$-~ algebra  generated by these density
functions
by using  the Gel'fand-Naimark theorem \cite{gel2}. In view of this fundamental
fact,
{\em we will hereafter assume
that in a quantum system, the classical configuration space and its
topological attributes are to be inferred from the
density functions
associated with $D(H)$ in the manner indicated above}.

More formally, our assumption is that in quantum theory, we have an
operator-valued hermitean form $\psi^* \chi$ with standard properties
\cite{var} for $\psi, \chi \in D(H)$, the scalar product $(\psi,\chi)$
being $Tr~ \psi^*\chi$. [This form in particular is to be positive in the
sense that $\psi^*\psi$ is a non-negative operator which vanishes iff
$\psi=0$.]  It is the above density function in conventional quantum
theory. There it generates an abelian normed *-algebra ${\cal A}^{\prime}$, the
norm and * being operator norm and hermitean conjugation, the latter
inverting $\psi^* \chi$ to $\chi^* \psi$: $(\psi^* \chi)^*=\chi^* \psi$.
[It may not be abelian in unconventional quantum theories such as those
on topological lattices \cite{top,bal3}]. According to our scheme, the
reconstruction of $Q$ from ${\cal A}^{\prime}$ is achieved by first
completing ${\cal A}^{\prime}$ to a $C^*$-algebra $\cal A$, and then using
the Gel'fand -Naimark theorem. Once $Q$ has been found, we can of course
identify $\cal A$ with $C(Q)$.

The recovery of $Q$ as a manifold requires also a $C^{\infty}$-~ structure
on $Q$. This can be specified in algebraic language by giving a suitable
subalgebra ${\cal A}^{(\infty)}$ of $\cal A$
\cite{mal}, the $C^{\infty}$-~
structure of $Q$ being that one for which
${\cal A}^{(\infty)}$ consists
of $C^{\infty}$ functions.

There is a natural way to specify
${\cal A}^{(\infty)}$ too in our scheme: Let $D^{\infty}(H)$ be the
subspace of $D(H)$ which is transformed into itself by
arbitrary powers of
$H$: $H^N D^{\infty}(H) {\underline{\subset}} D^{\infty}(H),~N=1,2 \cdots$.
Then
${\cal A}^{(\infty)}$ is generated by $\psi^* \chi$ when $\psi$ and
$\chi$ run over $D^{\infty}(H)$.

 There is a clear physical meaning to
$D^{\infty}(H)$ in terms of ultraviolet cut-offs and smooth time evolution as
we shall now show.

 Let us assume for
simplicity that the spectrum $\{E_n\}$ of $H$ is entirely discrete, and
let $H |E_n \rangle = E_n |E_n \rangle$, $\langle E_m|E_n \rangle=
\delta_{mn}$. It is then evident that $|E_n \rangle \in D^{\infty}(H)$.
Also any state vector of the form $\sum_n c_n |E_n\rangle \in D(H)$ where
$\sum_m |c_n E_n^N|^2 < \infty$ for $N=0,1,2 \cdots$ belongs to
$D^{\infty}(H)$. This condition is met if $|c_n|$ goes to zero
exponentially fast as $n \rightarrow \infty$.

High energy components of state vectors in $D^{\infty}(H)$ are thus
heavily suppressed. A consequence is that time evolution of vectors in
$D^{\infty}(H)$ is  smooth in time, arbitrary time derivatives
$\sum_n c_n(-iE_n)^Ne^{-iE_nt}|E_n\rangle$ of
$\sum_n c_ne^{-iE_nt}|E_n\rangle$ remaining in $D^{\infty}(H)$ if
$\sum_n c_n|E_n\rangle \in D^{\infty}(H)$.

In contrast, time evolution of vectors $\sum \, d_n\,|E_n \rangle \in
D(H)$ is generically only once-differentiable in time. Thus using the fact
that $H D(H) \in {\cal H}$, we can see that $\frac{d}{dt}\, \sum \,
d_n\,e^{-i E_n t}|E_n \rangle = \sum \, d_n\,(-i E_n)e^{-i E_n t}|E_n
\rangle \in {\cal H}$, but can not prove further differentiability of this
vector.

We find it striking that topological properties of the classical
configuration space $Q$ depend in this matter on the degree of temporal
smoothness. Vectors in $D(H)$ which are $C^1$ in time determine $Q$ as a
topological space whereas vectors in $D(H)$ which are $C^{\infty}$ in time
determine $Q$ as a manifold.

\sxn{A Simple Model}\label{se:3}

We will be considering particle dynamics from here onwards until Section
6. The configuration space of a particle being ordinary space, we are thus
imagining a physicist probing spatial topology using a particle.

Let us consider a particle with no internal degrees of freedom living on
the union $Q^{\prime}$ of two intervals which are numbered as 1 and 2:
\be
Q^{\prime}=[0,2\pi] \bigcup [0,2\pi] \equiv Q^{\prime}_1 \bigcup
Q^{\prime}_2~.\label{3.1}
\ee
It is convenient to write its wave function $\psi$ as $(\psi_1, \psi_2)$,
where each $\psi_i$ is a function on $[0,2\pi]$ and $\psi^*_i \psi_i$ is
the probability density on $Q^{\prime}_i$. The scalar product between
$\psi$ and another wave function $\chi=(\chi_1,\chi_2)$ is
\be
(\psi,\chi)=\int_0^{2 \pi} dx \sum_i (\psi^*_i\chi_i)(x)~.
\label{3.2}
\ee

It is interesting that we can also think of this particle as moving on
$[0,2\pi]$ and having an internal degree of freedom associated with the
index $i$.

After a convenient choice of units, we define the Hamiltonian formally by
\be
(H \psi)_i(x)=-\frac{d^2 \psi_i}{dx^2}(x)~ \label{3.3}
\ee
[where $\psi_i$ is assumed to be suitably differentiable in the interval
$[0,2\pi]$]. This definition is only formal as we must also specify its domain
$D(H)$
\cite{bal5}. The latter involves the statement of the boundary conditions
(BC's) at $x=0$ and $x=2\pi$.

Arbitrary BC's are not suitable to specify a domain: A symmetric operator
${\cal O}$
with domain $D({\cal O})$ will not be self-adjoint unless the following
criterion is also fulfilled:
\be
{\cal B}_{\cal O}(\psi,\chi)\equiv (\psi,{\cal O} \chi)-(
{\cal O}^{\dagger}\psi,\chi)=0~~{\it for~all}~\chi \in D({\cal O})
 \Leftrightarrow
\psi \in D({\cal O})~.
\label{3.4}
\ee

For the differential operator $H$, the form ${\cal B}_H( \cdot, \cdot)$ is
given by
\be
{\cal B}_H(\psi,\chi)
=\sum_{i=1}^2 \left[ -\psi^*_i(x) \frac{d
\chi_i(x)}{dx}+
\frac{d\psi^*_i(x)}{dx}\chi_i(x) \right]^{2 \pi}_0~.\label{3.5}
\ee
It is not difficult to show that there is a $U(4)$ worth of $D(H)$ here
compatible with (3.4).

We would like to restrict this enormous choice for $D(H)$, our intention
not being to study all possible domains for $D(H)$. So let us assume that
the momentum $P$ too is an observable and find possible $D(H)$
accordingly. That is to say, let us also insist that eigenstates of $P$
[and wave packets constructed as previously from its generalized eigenvectors
\cite{gel}, if any] belong to $D(H)$. We will achieve this end by
finding a domain $D(P)$ for $P$ and verifying that eigenvectors of this
$P$ are all in $D(H)$ when its BC's are properly chosen.

The momentum $P$ is defined by
\be
(P\psi)_i(x)=-i \frac{d \psi_i(x)}{dx}~.\label{3.6}
\ee
Hence
\be
{\cal B}_P(\psi,\chi)=-i \sum_{i=1}^2 \left[ \psi^*_i(x)
\chi_i(x)\right]^{2 \pi}_0~. \label{3.7}
\ee

Let $u\in U(2)$ [regarded as $2\times 2$ unitary matrices] and set
\be
D_u(P)= \left\{ \psi: \psi_i(2\pi)=u_{ij}\psi_j(0) \right\}~.\label{3.8}
\ee
In addition, $\psi$ must of course be differentiable.

It is now easy to verify that $P$ is self-adjoint for the domain $D_u(P)$
using the criterion (3.4). The eigenstates and spectrum of $P$ are
obtained by solving
\be
P\psi=p\psi,~~\psi\in D_u(P),~~~ p \in \bf{R}~.\label{3.9}
\ee
There is only discrete spectrum. The solutions are given by exponentials,
they are $C^{\infty}$ and are obviously in the domain
\be
D_u(H)= \left\{ \psi \in C^2(Q^{\prime}) :
\psi_i(2\pi)=u_{ij}\psi_j(0),~~~
 \frac{d\psi_i}{dx}(2\pi)=u_{ij}\frac{d\psi_j}{dx}(0)
\right\}~\label{3.10}
\ee
for $H$. Furthermore $H$ is (essentially) self-adjoint on this domain as
shown using (3.4). We therefore restrict attention to $D_u(H)$.

There are two choices of $u$ which are of particular interest:
\be
a)~~~~~~~~~~
 u_{a}=\left( \begin{array}{cc}
0 & e^{i\theta_{12}} \\
e^{i\theta_{21}} & 0
\end{array}\right) , \label{3.11}
\ee

\be
b)~~~~~~~~~~
 u_{b}=\left( \begin{array}{cc}
e^{i\theta_{11}} & 0 \\
0 & e^{i\theta_{22}}
\end{array} \right) .\label{3.12}
\ee

In case $a$, the density functions $\psi^*_i \chi_i$ fulfill
$$
(\psi^*_1 \chi_1)(2\pi)=(\psi^*_2 \chi_2)(0)~,
$$
\be
(\psi^*_2 \chi_2)(2 \pi)=(\psi^*_1 \chi_1)(0)~.\label{3.13}
\ee
Figure 1 displays (\ref{3.13}), these densities being the same at the points
connected by broken lines.

  \begin{figure}[hbtp]
  \epsfysize=4cm
  \epsfxsize=6cm
 \centerline{\epsfbox{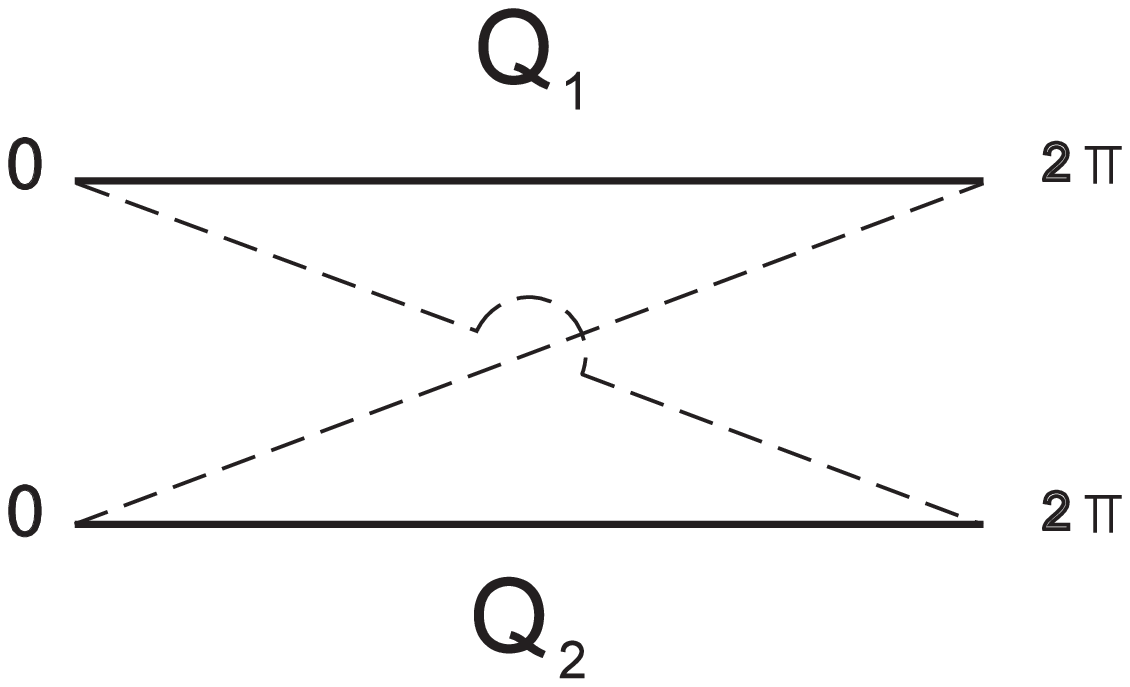}}
 \caption{  In case $a)$, the density
functions are the same at the points joined by broken lines in this Figure.}
\end{figure}

In case $b$, they fulfill, instead,
$$
(\psi^*_1 \chi_1)(2\pi)=(\psi^*_1 \chi_1)(0)~,
$$
\be
(\psi^*_2 \chi_2)(2 \pi)=(\psi^*_2 \chi_2)(0)~\label{3.14}
\ee
which fact is shown in a similar way in Figure 2.

\begin{figure}[hbtp]
  \epsfysize=4cm
  \epsfxsize=6cm
 \centerline{\epsfbox{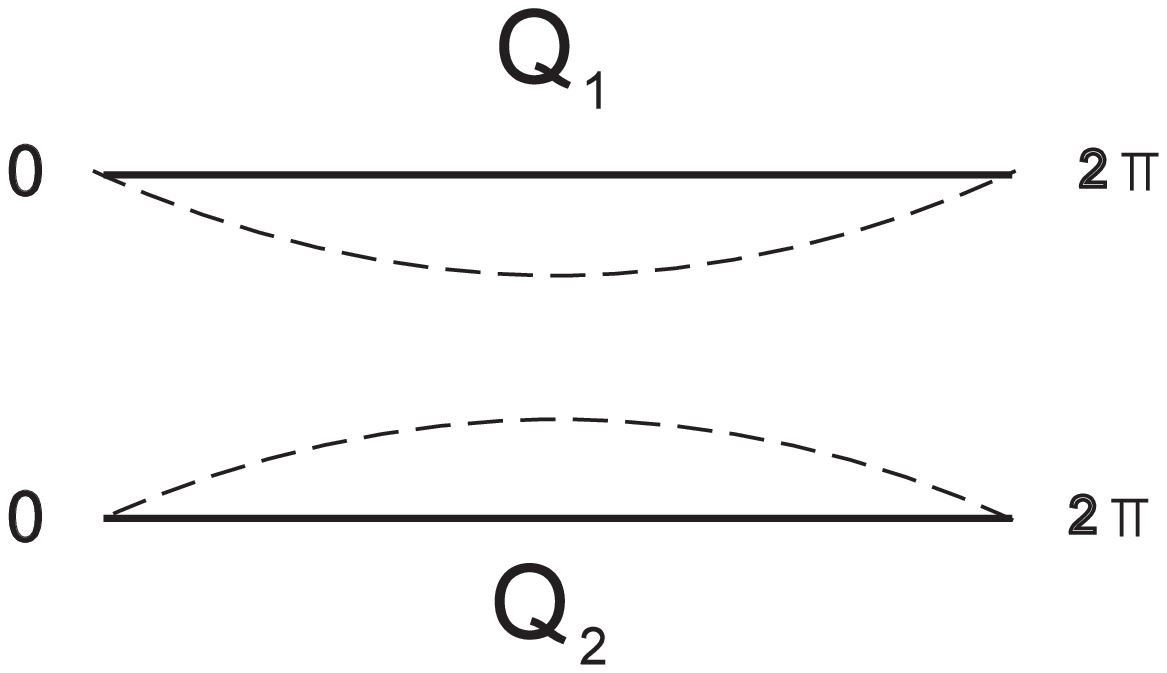}}
 \caption{ In case $b)$, the density
functions are the same at the points joined by broken lines in this Figure.}
\end{figure}

Continuity properties of $\psi^*_i \chi_i$ imply that we can identify the
points joined by dots to get the classical configuration space $Q$. It is
\underline{not} $Q^{\prime}$, but rather a circle $S^1$ in case $a$ and
the union $S^1 \bigcup S^1$ of two circles in case $b$.

The requirement $H^N D^{\infty}(H) \subset D^{\infty}(H)$ means just that
arbitrary derivatives of $\psi^*_i \chi_i$ are continuous at the points
joined by broken lines, that is on $S^1$ and $S^1 \bigcup S^1$ for the two
cases.
We can prove this easily using (\ref{3.10}). Thus on regarding $S^1$ and
$S^1\bigcup S^1$ as the usual manifolds, $D^{\infty}(H)$ becomes $
C^{\infty}(Q)$. In this way we also recover $Q$ as manifolds.

When $u$ has neither of the values (\ref{3.11}) and (\ref{3.12}), then $Q$
becomes the union of two intervals. The latter
happens for example for
\be
u=\frac{1}{\sqrt 2}\left( \begin{array}{cc}
1 & 1 \\
-1 & 1
\end{array}\right)~~. \label{3.15}
\ee
In all such cases, $Q$ can be regarded as a manifold with boundaries as shown
by the argument above.

Summarizing, we see that the character of the underlying classical
manifold depends on the domain $D_u(H)$ of the quantum Hamiltonian and
can change when $u$ is changed.

It is possible to reduce the $u$ in the BC to $\bf 1$ by introducing a
connection. Thus since $U(2)$ is connected, we can find a $V(x)\in U(2)$
such that
\be
V(0)={\bf 1},~~~~V(2 \pi)=u^{-1}.\label{3.16}
\ee
Using this $V$ we can unitarily transform $H$ to the new Hamiltonian
$$
H^{\prime}=VHV^{-1},
$$
$$
(H^{\prime}\psi)_i(x)=-\left[\frac{d}{dx}+A(x)\right]_{ij}^{2}\psi_j(x)~,
$$
\be
A(x)=V(x)\frac{d}{dx}V^{-1}(x)~.  \label{3.17}
\ee
With suitable physical interpretation, the system defined by $H^{\prime}$
and the domain
\be
D_{\bf 1}(H^{\prime})=V D_u(H)\equiv \{ \phi:~\phi=V\psi,~~\psi\in D_u(H)\}
\label{3.18}
\ee
is evidently equivalent to the system with Hamiltonian $H$ and domain
$D_u(H)$. Note in this connection that density functions on $Q_i^{\prime}$
are $\psi_i^*\chi_i$ and not $(V\psi)_i^*(V\chi)_i$.

\sxn{Dynamics for Boundary Conditions}\label{se:4}

We saw in the previous section that topology change can be achieved in
quantum physics by treating the parameters in the BC's as suitable
external parameters which can be varied. Here we
point out that there is no need for these parameters to be ``external" as they
too can
be treated as quantum variables.

Dynamics for $u$ which determines BC's is best introduced in the
connection picture where the domain of $H^{\prime}$ is associated with
$u=1$. We assume this representation hereafter.

Quantisation of $u$ is achieved as follows. Let $T(\alpha)$ be the
antihermitean generators of the Lie algebra of $U(2)$ [the latter being
regarded as the group of
$2
\times 2$ unitary matrices] and normalized according to $Tr~ T(\alpha)
T(\beta)=-N \delta_{\alpha \beta}$, $N$ being a constant. Let $\hat u$ be the
matrix of quantum operators representing the classical $u$. It fulfills

\be
\hat u_{ij} \hat u^{\dagger}_{ik} = {\bf 1} \delta_{jk},~~[\hat u_{ij}, \hat
u_{kh}]=0~, \label{4.1}
\ee
$\hat u_{ik}^{\dagger}$ being the adjoint of $\hat u_{ik}$.
The  operators conjugate to $\hat u$ will be denoted by $L_{\alpha}$. If
$$
[T_{\alpha},T_{\beta}]=c_{\alpha \beta}^{\gamma}T_{\gamma},
$$
\be
c_{\alpha \beta}^{\gamma}={\rm structure~constants~of}~U(2) \in {\bf R},
\label{4.2}
\ee
$L_{\alpha}$ has the commutators
$$
[L_{\alpha},\hat u]=-T(\alpha)\hat u~,
$$
\be
[L_{\alpha},L_{\beta}]=c_{\alpha\beta}^{\gamma}L_{\gamma}, \label{4.3}
\ee
$$
[T(\alpha)\hat u]_{ij}\equiv T(\alpha)_{ik}\hat u_{kj}.
$$

If $\hat V$ is the quantum operator for $V$, $[L_{\alpha},\hat V]$ is
determined by (\ref{3.17}) and (\ref{4.3}), $V$ being a function of $u$.

The Hamiltonian for the combined particle-$u$ system can be taken to be,
for example,
$$
\hat H = \hat H^{\prime} + \frac{1}{2I}\sum_{\alpha} L^2_{\alpha}
$$
\be
\hat H ^{\prime}=-\left( \frac{d}{dx}+ \hat A (x)\right)^2,~~\hat
A(x)\equiv \hat V(x) \frac{d}{dx} \hat V^{-1}(x), \label{4.4}
\ee
$I$ being the moment of inertia.

Quantised BC's with a particular dynamics are described by (\ref{4.1}),
(\ref{4.3}) and (\ref{4.4}).

The general wave function in the domain of $\hat H$ is a superposition of
state vectors $\phi \otimes_{\bf C}|u\rangle $ where $\phi \in D_{\bf
1}(H^{\prime})$ and $|u\rangle$ is a generalized eigenstate of $\hat u$:
\be
\hat u_{ij}|u\rangle=u_{ij}|u\rangle,~~\langle u^{\prime}|u
\rangle=\delta(u^{\prime -1}u)~.\label{4.5}
\ee
The $\delta$-function here is defined by
\be
\int du f(u)\delta(u^{\prime -1}u)=f(u^{\prime}),\label{4.6}
\ee
$du$ being the (conveniently normalized) Haar measure on $U(2)$. Also
\be
\hat A(x) |u\rangle = A(x) |u \rangle~.\label{4.7}
\ee

It follows that the classical topology of one and two circles is
recovered on the states $\sum C_{\lambda}
\phi^{(\lambda)}\otimes_{\bf C}|u_{a}\rangle $ and
$\sum
D_{\lambda}\phi^{(\lambda)}\otimes_{\bf
C}|u_b\rangle,~[C_{\lambda},~D_{\lambda}\in
{\bf C}, ~\phi^{(\lambda)}\in D_{\bf 1}(H^{\prime})$] with the two fixed
values $u_a$,  and $u_b$ of (\ref{3.11}) and (\ref{3.12}) for $u$.

But these are clearly idealized unphysical vectors with infinite norm.
The best we can do with normalizable vectors to localize topology around
one or two circles is to work with wave packets

$$
\int du f(u) \phi \otimes_{\bf C} |u\rangle~,
$$
\be
\int du |f(u)|^2 < \infty \label{4.8}
\ee
where $f$ is sharply peaked at the $u$ for the desired topology. The
classical topology recovered from these states will only approximately be
one or two circles, the quantum topology also containing admixtures from
neighbouring topologies of  two intervals.

A localised state vector of the form (\ref{4.8}) is not as a rule an
eigenstate of a Hamiltonian like $\hat H$. Rather it will spread in
course of time so that classical topology is likely to
 disintegrate
mostly into that of two intervals.  We can of
course localise it around one or two $S^1$'s for a very long time by
choosing $I$ to be large, the classical limit for topology being achieved
by letting $I \rightarrow \infty$. By adding suitable potential terms, we
can also no doubt arrange matters so that a wave packet concentrated
around $u=u_a$ moves in time to one concentrated around  $u=u_b$. This
process would be thought of as topology change by a classical physicist.

\sxn{Generalizations}\label{se:5}

Considerations of the last sections admit generalizations to higher
dimensions which we now indicate.

In analogy with the previous two-interval model, we now start our
discussion with
\be
Q^{\prime}=C_1\bigcup C_2~.\label{5.1}
\ee
Here $C_i$ are two cylinders. We assume for convenience that
they are identical and can thus be identified with a common cylinder $C$:
\be
C= \{(x_1,x_2)~:~
x_1 \in [0,2 \pi],~~x_2 \in [0,2 \pi]\}. \label{5.2}
\ee
Here $(0,x_2)$ and $(2\pi,x_2)$ are to be identified.

Let us consider a particle on $Q'$ with spin associated with a two-valued
index. The wave function $\psi$ can then be written as $(\psi_1,\psi_2)$
where each $\psi_i$ has two components:
$\psi_i=(\psi_i^{(1)},\psi_i^{(2)})$. Here $\psi_i^{(\rho)}$ is a ${\bf
C}$-valued function on $C_i$ and $\psi_i ^{\dagger}\psi_i$ is
its
probability density. As for the scalar product and Hamiltonian, we choose
them to be
$$
(\psi,\chi)=
\int_{C} dx_1 \, dx_2 \,\sum_i \psi^{\dagger}_i\chi_i(x)
\equiv
\int_{C} dx_1 \, dx_2 \,\psi^{\dagger}\chi(x)~,
$$
\be
(H \psi)^{(\rho)}_i(x)=-\sum_{j=1}^2\frac{\partial^2
\psi_i^{(\rho)}}{\partial x_j^2}(x)~,~~x=x_1,x_2~. \label{5.3}
\ee

Let
$$
{\cal D}= -i\sum_{i=1}^2 \alpha_i \partial_i,
$$
\be
\alpha_1=\tau_1,~~\alpha_2=\tau_3,~~~~\tau_i={\rm Pauli~matrices}
\label{5.4}
\ee
be the  Dirac operator. As ${\cal D}^2=H$, we will use ${\cal D}$ as the
substitute for the momentum operator of Section 3 to simplify considerations.

Now, we have
$$
{\cal B}_{\cal D}(\psi,\chi)=(\psi,{\cal D}\chi)-({\cal D}^{\dagger}
\psi,\chi)= $$
\be
=
i\int_{x_2=0} dx_1 \,~\psi^{\dagger}\alpha_2 \chi-
i\int_{x_2=2 \pi} dx_1 \,~\psi^{\dagger}\alpha_2 \chi~. \label{5.5}
\ee
We will establish the feasibility of topology change even in this
two-dimensional example, limiting ourselves for simplicity to a particular
class of
BC's for which $\cal D$ is (essentially) self-adjoint. They are labelled by an
element
of $U(2)$ [regarded as $2 \times 2$ unitary matrices] and are given by
$$
D_u({\cal D})=\{ \psi:~\psi_i^{(k)}\in {\cal
C}^1(Q^{\prime});
$$
\be
\psi^{(k)}_i(x_1, 2 \pi)=u_{ij}
\psi^{(k)}_j(x_1,0)\}~.\label{5.6}
\ee

As for $H$, we can make it (essentially) self-adjoint on a dense
subset $D_u(H)$ of $D_u(\cal D)$
obtained by imposing also twice-differentiability and additional conditions:
$$
D_u(H)=\{ \psi:~\psi_i^{(k)}\in C^2(Q^{\prime})~;
$$

$$
\psi^{(k)}_i(x_1, 2 \pi)=u_{ij}
\psi^{(k)}_j(x_1,0),
$$

$$
\partial_{x_2}\psi^{(k)}_i(x_1, 2 \pi)=u_{ij}\partial_{x_2}
\psi^{(k)}_j(x_1,0)~\}~;
$$

\be
\partial_{x_2}\equiv \frac {\partial}{\partial x_{2}}~. \label{5.7}
\ee

It is easily seen that for the choice $u=u_a$ [eq. (\ref{3.11})],
the density functions remain continuous if we identify the points of the
boundaries of $C_1$ and $C_2$ as shown in Figure 3. The result is a
single torus.
Alternatively we can also choose $u=u_b$ [eq. (\ref{3.12})].
With this latter $u$, $Q$ becomes the union of two tori. For other choices
of $u$ in eq. (\ref{5.6}), $Q$ becomes the union of two
cylinders.

More general topologies for $Q$ can be obtained by allowing the matrix $u$
in eq. (\ref{5.7}) to depend on the coordinate $x_1$. The choice
\be
u(x_1)=\left\{ \begin{array}{ll} {\bf 1} & \mbox{for $x_1 \in
[0,\frac{\pi}{2}] \bigcup [\frac{3 \pi}{2}, 2 \pi]$} \\
\tau_1 & \mbox{for $x_1\in[\frac{\pi}{2}, \frac{3 \pi}{2} ]$}
\end{array} \right. \label{5.8}
\ee
for example makes $Q$ a genus two surface as shown in Figure 3.

\begin{figure}[hbtp]
  \epsfysize=5cm
  \epsfxsize=13cm
 \centerline{\epsfbox{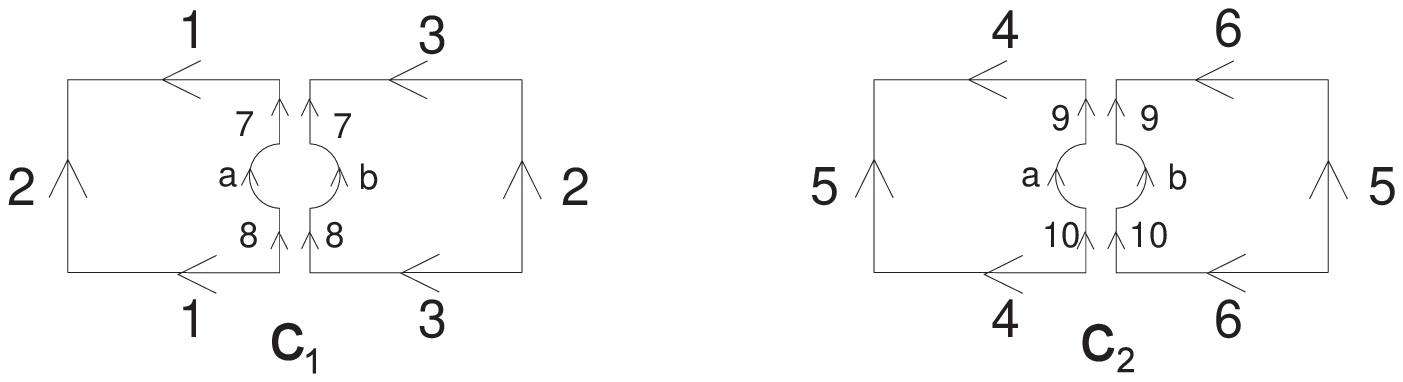}}
 \caption{  Opposite sides on the periphery with the same labels
are to be
identified in the direction of the arrows. With the cuts in the middle,
$C_1$ and $C_2$ are thus cylinders. On identifying the lips of the cuts
in the direction of the arrows in each cylinder separately, we get two
tori. If instead we identify the lips of the cuts with the same labels
in the direction of the arrows, we get a genus two manifold. If we do this
identification for all except $a$ and $b$, $C_1$ and $C_2$ become tori
with holes. }
\end{figure}

There is the following simple way to see that (5.8) gives a genus two surface.
With just the first line of (5.8), the effective configuration space is the
union of two tori, each with a hole.  These holes get identified with the
second line of (5.8) to give us the genus two surface.
This process is also shown in Figure 3.

We could in fact have begun with a pair of tori with holes and then used the
second line  of (5.8) as the BC.  In this approach, we see clearly that what is
involved is taking connected sums [1-3] of tori.  It also becomes obvious that
we can in this way pass from manifolds with genera  $g_1$ and $g_2$ to get a
manifold with  genus $g_1+g_2$.

It is evident that topology change can be achieved here by changing $u$. For
example a genus $g$ manifold can be split into manifolds with genera  $g_1$ and
$g_2$, $g$ being $g_1+g_2$.  It should also be clear that much of the work in
the previous sections can be adapted to such models.

There seems also to be no barrier to higher dimensional generalisations.

It is interesting that the process whereby  surfaces are joined together in our
approach to obtain a closed (compact and boundaryless) manifold is
{\em local},~ in that the identified set contains at most one point from each
cut.  But this is enough both for connected sums [1-3], and, in three
dimensions, for
surgery  on links [19]. As any orientable, closed
 three-manifold can be changed to any other such manifold by surgery on
links [19], we speculate that corresponding topology changes can also be
achieved by
our methods. There seems also to be no problem in taking connected sums
[1-3]
of these manifolds with $\bf{R}^2$ or $\bf{R}^3$ and extending these
considerations to asymptotically flat spatial slices.

\sxn{Final Remarks}\label{se:6}

We have seen in this article that topology change can be achieved in
quantum physics by judicious introduction of new degrees of freedom.
They  control the BC's of operators associated with the classical
configuration space. When they  change, the classical configuration space  too
is  changed
and suffers topological transmutations.

There is striking resemblance of this mechanism for topology change and the
axion approach to the strong CP problem in QCD \cite{axi}. The latter is
caused by the fact that its Hamiltonian admits a one-parameter family of
boundary conditions on its wave functions, labelled by an element
$e^{i\theta}$ of $U(1)$. It has thus a one-parameter family of domains.
The possibility of these BC's is reflected by the $\theta Tr(F \wedge F)$
term in the action.

If the BC's are now made dynamical, $\theta$ becomes the axion field $a$
and the above term in the action becomes $aTr(F\wedge F)$.

In this model with the additional axion degree of freedom, we find the
$\theta$-vacuum when $a$ is frozen to the value $\theta$. But $a$ will in
general fluctuate between different values so that the effective QCD
$\theta$ depends on the state of $a$. These features resemble those found
in the previous sections.

In any case, we see that topology change is facilitated by introducing
new degrees of freedom. Let us  now also recall the existence of indications
that
topology change should be permitted in quantum gravity to enforce the
standard spin-statistics connection. The reasons leading to this opinion
were summarized in the very first section.

The above two aspects relating to topology change suggest that {\em any
eventual theory of quantum gravity will contain degrees of freedom going
beyond those in Einstein gravity or its minor variants.}

There is another line of thought indicating that Einstein's model for
gravity is somehow only an effective theory of another underlying theory
with more degrees of freedom. It has got repeatedly emphasised during
conversations with Rafael Sorkin over the years and runs as follows. In
molecular physics in the Born-Oppenheimer approximation, or in the
collective model for nuclei, molecules and nuclei are generally described as
rigid
bodies with discrete symmetry groups $G \subset SU(2)$. These groups are also
the
fundamental groups $\pi_1(Q)$ of their configuration spaces $Q$, which on
ignoring translations, are $SU(2)/G$ [20,3]. When $\pi_1(Q)$ is
nontrivial, there are as many ways of quantising the system as there are
unitary
irreducible representations (UIRR's) of $G$ \cite{bal1}. These
uncertainties correspond to uncertainties in the choice of BC's for the
wave functions or equivalently the domain of the Hamiltonian. The
distinct quantisations can be so different that wave functions are
spinorial in one and tensorial in another [3,20].

Now, it is often the case that molecules and nuclei described by these
different UIRR's do occur in nature. In their microscopic structure, they
differ in their nuclear and/or electronic constituents. One can say that
quantisation ambiguities in the Born-Oppenheimer approximation or in the
collective nuclear models reflect the possibility of differing
microscopic constituents for the same rigid body. Indeed the particular
quantisation appropriate for a molecular or nuclear species is chosen  by
chemists and nuclear physicists by appealing to its microscopic description.

In a similar way, the fundamental group $\pi_1(Q)$ of the configuration
space $Q$ for the two flavour Skyrme model is ${\bf Z}_2$ \cite{bal1}. It
therefore has a two-fold uncertainty of quantisation  leading
respectively to a spinorial and a tensorial soliton \cite{bal1}. Here too
this ambiguity reflects the possibility of differing microscopic
constituents for solitons and can be resolved by postulating a specific
microscopic structure. This microstructure is provided by QCD. The
soliton is spinorial if the number of colours is odd, and tensorial if it
is even.

These examples suggest that quantisation ambiguities may indicate an
underlying microstructure and its associated novel degrees of freedom.

Let us now turn to gravity. Here the fundamental group $\pi_1(Q)$ of the
configuration space $Q$ is extremely complex in the presence of geons
[21,1-3] leading to enormous quantisation uncertainties.
Our experience in molecular, nuclear and particle physics now suggests
strongly that {\em gravity too has an underlying microscopic structure with
its novel degrees of freedom, and that these ambiguities merely
reflect the fact that it is an effective theory of several differing
microscopic theories.}

The preceeding arguments, one based on the spin-statistics connection and
the second on quantisation ambiguities, lend encouragement to radical
attempts at deriving Einstein gravity [22] and even spacetime
as a manifold \cite{bom} from deeper microscopic models.

\section{Acknowledgements}

The ideas of this paper relating to gravity have been strongly influenced
by Rafael Sorkin. Some of them originate from him or in
conversations with him. We thank him for numerous discussions over the
years. We also thank Giovanni Sparano and the coauthors of two of us in
refs. \cite{bal3} and \cite{bal4} for discussions and for contributions to
aspects of this paper. Finally we are very grateful to Arshad Momen and Rafael
Sorkin with much
help with the references,and L.Chandar, Arshad Momen, Carl Rosenzweig,
 Sumati Surya and Sachin Vaidya for valuable suggestions regarding the
manuscript. This work was supported by the Department of
Energy under contract number DE-FG-02-85ER40231.

%
%


\begin{thebibliography}{99}

\bibitem{fri} J.L. Friedman and R.D. Sorkin, {\it Phys. Rev. Lett.}  {\bf 44}
(1990) 1100; ibid. {\bf 45} (1980) 148;
{\it Gen. Rel. Grav.} {\bf 14} (1982) 615.

\bibitem{sor} R.D. Sorkin; {\it Introduction to Topological Geons},
in {\it Topological Properties and Global Structure of Space-Time}, eds. P.G.
Bergman and V. de Sabata (Plenum, 1986); {\it Classical Topology
and Quantum
Phases: Quantum Geons}, in  {\it Geometrical and Algebraic
Aspects of Nonlinear
Field Theory}, eds. S. De Filippo, M. Marinaro, G. Marmo and  G. Vilasi
(North-Holland, 1989).

\bibitem{bal1} A. P. Balachandran, G. Marmo, B. S. Skagerstam and A.
Stern, {\it Classical Topology and Quantum States} [World Scientific, 1991].


\bibitem{bal2} A.P. Balachandran, A. Daughton, Z.-C. Gu, G. Marmo, R.D. Sorkin,
and
A.M. Srivastava, {\it Mod. Phys. Lett.} {\bf A5} (1990) 1575 and {\it
Int. J. Mod. Phys.} {\bf A8} (1993) 2993;  A.P. Balachandran, W.D. McGlinn, L.
O'Raifeartaigh, S. Sen and R.D. Sorkin, {\it Int. J. Mod. Phys.} {\bf A7}
(1992) 6887 [Errata: {\it Int. J. Mod. Phys.} {\bf A9} (1994) 1395]; A.P.
Balachandran, W.D. McGlinn, L. O'Raifeartaigh, S. Sen, R.D. Sorkin and A.M.
Srivastava, {\it Mod. Phys. Lett.} {\bf A7} (1992) 1427.



\bibitem{dow}F. Dowker and R.D. Sorkin, {\it A Spin-Statistics
Theorem for Certain
Topological Geons}, Fermi Lab preprint FERMILAB-Pub-93/278-A (1993).


\bibitem{see} See especially refs. 2 and 5 in this regard.

\bibitem{bom} L. Bombelli, J. Lee, D. Meyer and R.D. Sorkin, {\it Phys. Rev.
Lett.} {\bf 59} (1987) 521;
R.D. Sorkin, {\it Int. J. Theor. Phys.} {\bf 30} (1991) 923.

\bibitem{bal3}
A.P. Balachandran, G. Bimonte, E. Ercolessi, G.
Landi, F. Lizzi, G. Sparano and P. Teotonio-Sobrinho, {\it
Finite Quantum Physics and Noncommutative Geometry} [IC/94/38, DSF-T-2/94
SU-4240-567 (1994), hep-th/9403067], in {\it Proceedings of the XV Autumn
School,`` Particle Physics in the 90's'' }, eds. G. Branco and G. Pimenta,
{\it Nucl. Phys. B} ({\em Proc. Suppl.}) {\bf 37C} (1995) 20; {\it J. Geom.
Phys.} (in press).

\bibitem{bal4} A.P. Balachandran and L. Chandar, {\it Nucl. Phys.} {\bf B428}
(1994)435.

\bibitem{bal5} M. Reed and B. Simon,{\it Methods of Mathematical Physics},
Vol. II:
{\it Fourier Analysis; Self Adjointness} [Academic Press, 1975].

\bibitem{imb} A recent striking work with close relation to this paper is
that of T. Imbo and P. Teotonio-Sobrinho, University of Illinois, Chicago
preprint (in preparation).

\bibitem{axi} For reviews of axion physics, see for example J.E. Kim, {\it
Phys. Rep.} {\bf 150} (1987) 1; H.Y. Cheng, {\it Phys. Rep.} {\bf 158} (1988)
1; M.S. Turner, {\it Phys. Rep.} {\bf 197} (1990) 67; P. Sikivie, {\it
Axion
Searches}
in  {\it Perspectives in the Standard Model, Proceedings of the 1991
Theoretical
Advanced Study Institute in Elementary Particle Physics}, Boulder, Colorado,
2-28 June 1991,  eds. R.K. Ellis, C.T. Hill and J.D. Lykken (World
Scientific, 1992).


\bibitem{rs} B. Simon and M. Reed, {\it Mathematical Methods of Modern
Physics}, Vol. I: (Academic Press, 1972).

\bibitem{gel2} J.M.G. Fell and R.S. Doran, {\it Representations of
$^*$-Algebras, Locally Compact Groups and Banach $^*$-Algebraic Bundles}
[Academic press, 1988].

\bibitem{var} J.C. V{\'a}rilly and J.M. Gracia-Bond{\'i}a, {\it
J. Geom. Phys.}
{\bf 12} (1993) 223.


\bibitem{top} A.P. Balachandran, G. Bimonte. E. Ercolessi and
P. Teotonio-Sobrinho, {\it Nucl. Phys.} {\bf B418} (1994) 477.

\bibitem{mal} A. Mallios, {\it Topological Algebras: Selected Topics},
Mathematics Studies 124 [North- Holland, 1986].

\bibitem{gel} I.M. Gel'fand and N.Ya. Vilenkin, {\it Generalized
Eigenfunctions}, Vol. 4: {\it Applications of Harmonic Analysis} [Academic
Press, 1964].

\bibitem{auf} D. Rolfsen, {\it Knots and Links} [Publish or Perish, 1976], Ch.
9; L.H.
Kauffman and S.L. Lins,{\it Temperley-Lieb Recoupling Theory and Invariants
of 3-Manifolds}, Annals of Mathematics Studies, Number 134
[Princeton University
Press, 1994], p. 134.

\bibitem{sim} A.P. Balachandran, A. Simoni and D.M. Witt, Int. J. Mod.
Phys. {\bf A7} (1992) 2087.

\bibitem{wit} D.M. Witt, J. Math. Phys. {\bf 27} (1986) 573.

\bibitem{adl}    Some papers discussing ``induced gravity'' using which further
references can be traced are Ya. B. Zel'dovich, {\it JETP Lett.} {\bf 6}
(1967) 316;
A.D. Sakharov, {\it Sov. Phys. Doklady} {\bf 12} (1968) 1040; A. Zee, {\it
Phys. Rev.} {\bf D23} (1981) 858; S.L. Adler, {\it Rev. Mod. Phys.} {\bf 54}
(1982) 729; D. Amati and G. Veneziano, {\it Nucl. Phys.} {\bf B204} (1982)
 451;
F. David, {\it Phys. Lett.} {\bf 138B} (1984) 383;  A. Strominger, {\it
Is there a
Quantum Theory of Gravity} and E. Tomboulis, {\it Renormalization and
Asymptotic
Freedom in Quantum Gravity,} in  {\it Quantum Theory of Gravity}, ed. S.M.
Christensen [A. Hilger, 1984]; T. Jacobson, University of Maryland preprint
(1994), gr-qc 9404039.



\end{thebibliography}
\end{document}